\documentclass[10pt]{article}
\usepackage{amsmath, amssymb, graphicx, setspace,natbib,hyperref}
\usepackage{genetics}

\newcommand{\eq}[1]{Eq.~(\ref{#1})}
\newcommand{\eqs}[2]{Eqs.~(\ref{#1}) and (\ref{#2})}
\newcommand{\ev}[1]{E\left[ #1 \right]}

\newcommand{\fig}[1]{Fig.~\ref{fig:#1}}

\newcommand{\bigo}[1]{\mathcal{O}\left(#1\right)}
\newcommand{\SI}[1]{\ref{si:#1}}

\newcommand{\rc}{r_\times}
\newcommand{\vc}{v_\times}
\newcommand{\evs}{\bar s}
\newcommand{\qc}{q_\times}
\newcommand{\oxf}{f}
\newcommand{\var}{\text{Var}}

\newcommand{\test}{t_\text{est}}
\newcommand{\tnm}{t_\text{nm}}
\newcommand{\nest}{n_\text{est}}

\begin{document}

\title{The rate of adaptation in large sexual populations with linear chromosomes}
\author{DB Weissman$^{1,3}$, O Hallatschek$^{2,3}$\\
$^1$\textit{Institute of Science and Technology Austria}\\
$^2$\textit{Department of Physics and } $^3$\textit{California Institute for Quantitative Biosciences,} \\
\textit{University of California, Berkeley, CA 94720}}
\date{}

\maketitle

\begin{abstract}
In large populations, multiple beneficial mutations may be simultaneously spreading.
In asexual populations, these mutations must either arise on the same background or compete against each other. 
In sexual populations, recombination can bring together beneficial alleles from different backgrounds, but tightly linked alleles may still greatly interfere with each other. 
We show for well-mixed populations that when this interference is strong, the genome can be seen as consisting of many effectively asexual stretches linked together.
The rate at which beneficial alleles fix is thus roughly proportional to the rate of recombination, 
and depends only logarithmically on the mutation supply and the strength of selection.
Our scaling arguments also allow to predict, with
  reasonable accuracy, the fitness distribution of fixed mutations when
  the mutational effect sizes are broad. 
  We focus on the regime in which crossovers occur more frequently than beneficial mutations, 
 as is likely to be the case for many natural populations.
  
\end{abstract}

\section*{Introduction}

In a large, adapting population, beneficial alleles may be simultaneously spreading at multiple loci.
These alleles will tend to arise in different lineages and compete with each other, slowing adaptation, an effect known as ``(clonal) interference.'' 
This phenomenon has been repeatedly observed in many different microbial and viral evolution experiments \citep{lenski1991, devisser1999, miralles1999, colegrave2002, goddard2005, hegreness2006, desai2007a, bollback2007, pepin2008, kao2008, barrick2009, betancourt2009, lang2011, miller2011};
recently, it has also been demonstrated to be occurring natural viral populations \citep{batorsky2011, strelkowa2012,ganusov2013}. 
Recombination alleviates interference by breaking down negative associations among the beneficial alleles;
in fact, it has long been thought that this effect may be a reason for the evolution of sexual reproduction \citep{weismann1889,fisher1930,muller1932}.
Although the effect of interference on the rate of adaptation in sexual populations has recently been the subject of a substantial amount of theoretical analysis
 \citep{cohen2005, cohen2006, rouzine2005, rouzine2007, rouzine2010, neher2010, batorsky2011, neher2011, weissman2012}, 
 this has mostly been restricted to considering interference among unlinked loci (i.e., with all loci reassorting independently at a uniform rate),
 or else limited to weak-to-moderate interference in which only rare alleles are affected.
 
 However, for many real populations, particularly viral ones, interference may be both strong and primarily occurring among tightly-linked loci,
 so that at any given time each polymorphic beneficial allele is simultaneously interacting with multiple other alleles at varying recombination fractions.
 In organisms such as viruses and eukaryotes in which recombination within chromosomes or genome segments occurs primarily via crossovers, 
these recombination fractions vary hugely among different pairs of loci. 
In humans, for example, loci at opposite ends of a chromosome are unlinked, with recombination rate $r=0.5$, while pairs of loci within a gene may have
recombination rates of less than $r\sim 10^{-6}$ \citep{myers2005}. 
Less is known about viral recombination rates, but \cite{neher2010a} estimate that in HIV recombination rates among loci vary by a factor of $\sim 10^3$ over the genome.
 In this paper, we conduct the first analysis of adaptation under strong interference in such populations with large ranges of recombination fractions among loci.
\cite{neher2013} conduct a very similar analysis for a complementary region of parameter space (see ``Model'' below). 

\section*{Model}

We consider a well-mixed population of $N$ haploid individuals. Individuals reproduce sexually, producing each offspring with a different mate. 
The genome consists of a single chromosome of map length $R$, i.e., there are an average of $R$ crossovers per reproduction. 
(All of our results apply equally well to a population of facultative sexuals that outcross with frequency $\oxf<1$, with $R$ as the map length multiplied by $\oxf$;
issues only begin to arise when $4v\gtrsim \oxf^2$ -- see \SI{loose}.)
There is a constant genomic beneficial mutation rate $U_b$, regardless of genetic background, so that beneficial mutations are never exhausted. 
We ignore deleterious mutations and epistasis among polymorphic beneficial mutations. 
With these assumptions, the population will approach an expected steady long-term rate of adaptive substitution per unit genetic map length,
$\lambda$, and the rate of increase in mean log fitness, $v$; we focus on populations close to this steady state.
It will be useful to consider the rate of beneficial mutation per unit map length, $\mu_b\equiv U_b/R$.
We will focus on the case in which beneficial mutations are infrequent relative to recombination ($\mu_b<1$); 
\cite{neher2013} consider the opposite case.

\subsection*{Approach}

Even in sexual populations, short stretches of genome will be effectively asexual -- 
all loci within a sufficiently short stretch are likely to be described by the same genealogy tracing back to a single common ancestor. 
In other words, short stretches will typically coalesce without undergoing any recombination. 
Since each selective sweep increases the rate of coalescence via genetic draft, in rapidly-adapting populations
these stretches may be long enough so that multiple beneficial alleles will be almost completely linked for the entire time that they are polymorphic,
and will therefore strongly interfere with each other.
On the other hand, once outside this stretch, the strength of interference decays rapidly, with a dependence on recombination rate approaching $\propto r^{-2}$ 
\citep{weissman2012}.

This combination of strong interference among very tightly linked alleles and weak interference among the rest of the genome
suggests approximating the genome as consisting of a series of effectively asexual stretches that have relatively little effect on each other. 
This allows us to avoid the difficulties that come from dealing with an explicit model of crossovers between linear genomes. 
To understand the evolution of the short stretches, we can draw on the extensive theoretical work on the evolution of asexual populations in the strong-interference
regime, particularly \cite{desai2007c, rouzine2008, good2012}.

\section*{Results}

\subsection*{Length of effectively asexual stretches, $\mathbf\rc$, and the rate of adaptation}

\begin{figure}
   \includegraphics[width=6in]{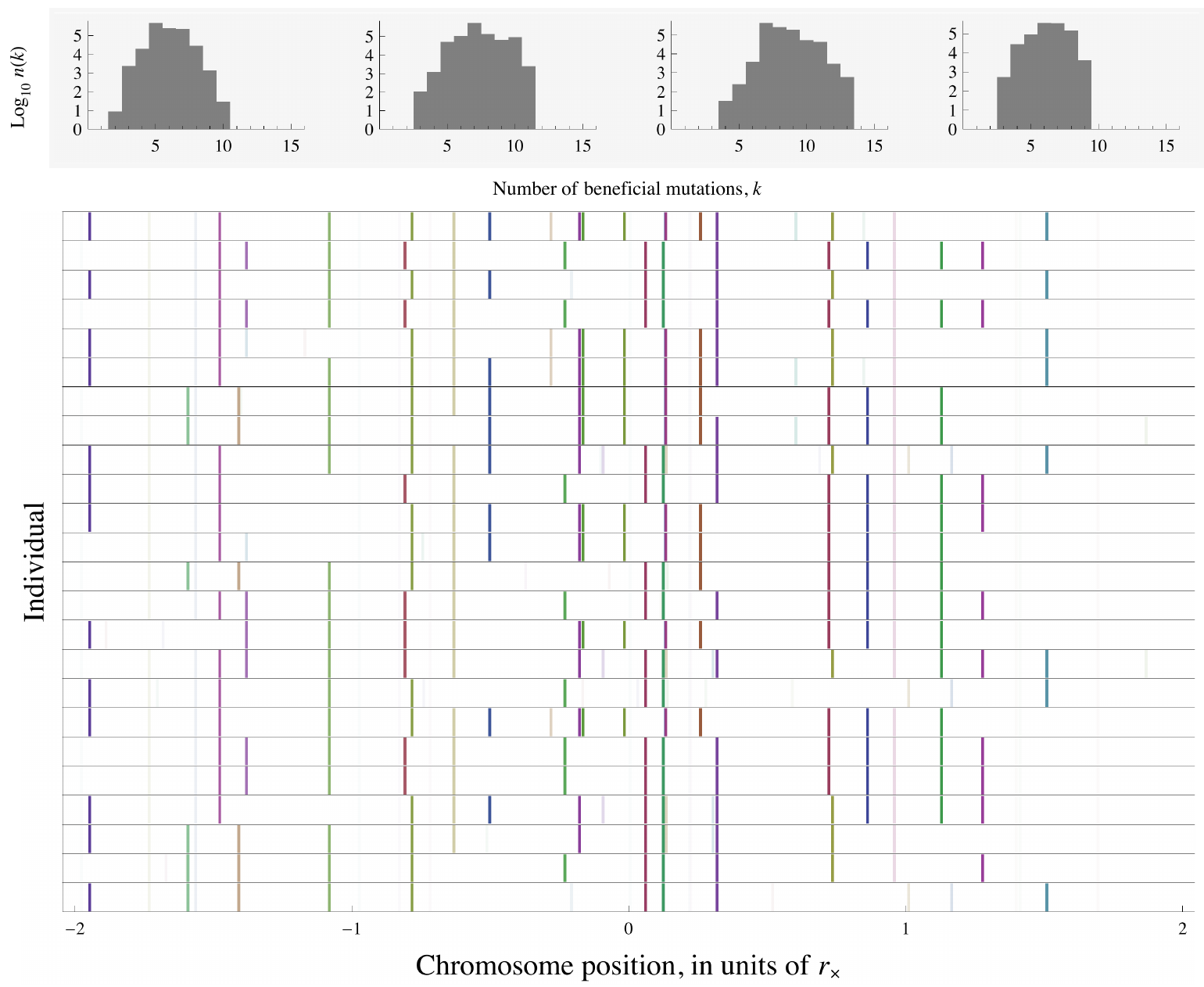}
   \caption{\label{fig:snapshot}
   Sample of a small region of the genome in 25 individuals from a simulated population. 
   Colored bars are beneficial mutations, with mutations at very low frequencies or approaching fixation shown in faded colors.
   Distance along the chromosome is measured in units of the predicted asexual  chunk  size, $\rc\approx0.01$ (\eq{rc}).
   At distances $r<\rc$, there is very strong linkage disequilibrium, which rapidly decays for $r>\rc$. 
  The histograms above the sample show the distributions of beneficial mutations in the chunk below over all individuals in the population.
  The distance from  the mean of the distribution to the nose is close to the predicted value, $\lambda\approx 4$ (\eq{lam}; see \SI{calc}).
  Simulation parameters are $N=10^6$, $s=0.05$, $\mu_b=0.1$.
   }
   \end{figure}

We want $\rc$ to be the largest genetic scale over which the population evolves approximately asexually. 
In other words, mutations that arise at nearby loci separated by recombination fraction $r<\rc$ 
should be strongly associated with each other (either positively or negatively) until they fix or go extinct,
while those at loci separated by $r>\rc$
should spread independently from each other, i.e., should be in linkage equilibrium by the time they become
common and might start to affect each other.
\fig{snapshot} shows this pattern in a sample from a simulated population.
Since new mutations start in strong linkage disequilibrium, which then decays at an average rate $r$, 
this is equivalent to requiring that beneficial alleles take a time $\sim 1/ \rc$ to become common starting from a single copy.

The time for a new mutation to reach high frequency itself depends on the amount of interference.
To find it, we will make the approximation that loci farther than $\rc$ from each other are essentially unlinked. 
Then we can treat the genome as consisting of $\sim R/\rc$ independently evolving asexual ``chunks,'' each with beneficial mutation rate $\approx \mu_b\rc$.
The evolution of each chunk can be described as a traveling wave in fitness space.
For the most part, only the beneficial mutations that arise near the nose of the wave have a chance of reaching high frequency,
and the time it take for them to do so is roughly the time for the wave to travel the distance between its mean and its nose.
Thus $\rc$ should be approximately equal to the speed of the fitness wave divided by its mean-to-nose width.
Equivalently, looking backwards in time, $\rc$ is set by the condition that the individual in the distant past carrying the ancestor of present-day allele should 
be very likely to have very high fitness at loci within $r\sim\rc$ of the focal allele but should have roughly average fitness elsewhere.
In a  traveling fitness wave, the time for a typical individual's ancestry to trace back to the nose
 is the mean-to-nose width divided by the speed of the wave \citep{desai2013},
giving the same value for $\rc$.

\begin{figure}
 \includegraphics[width=6in]{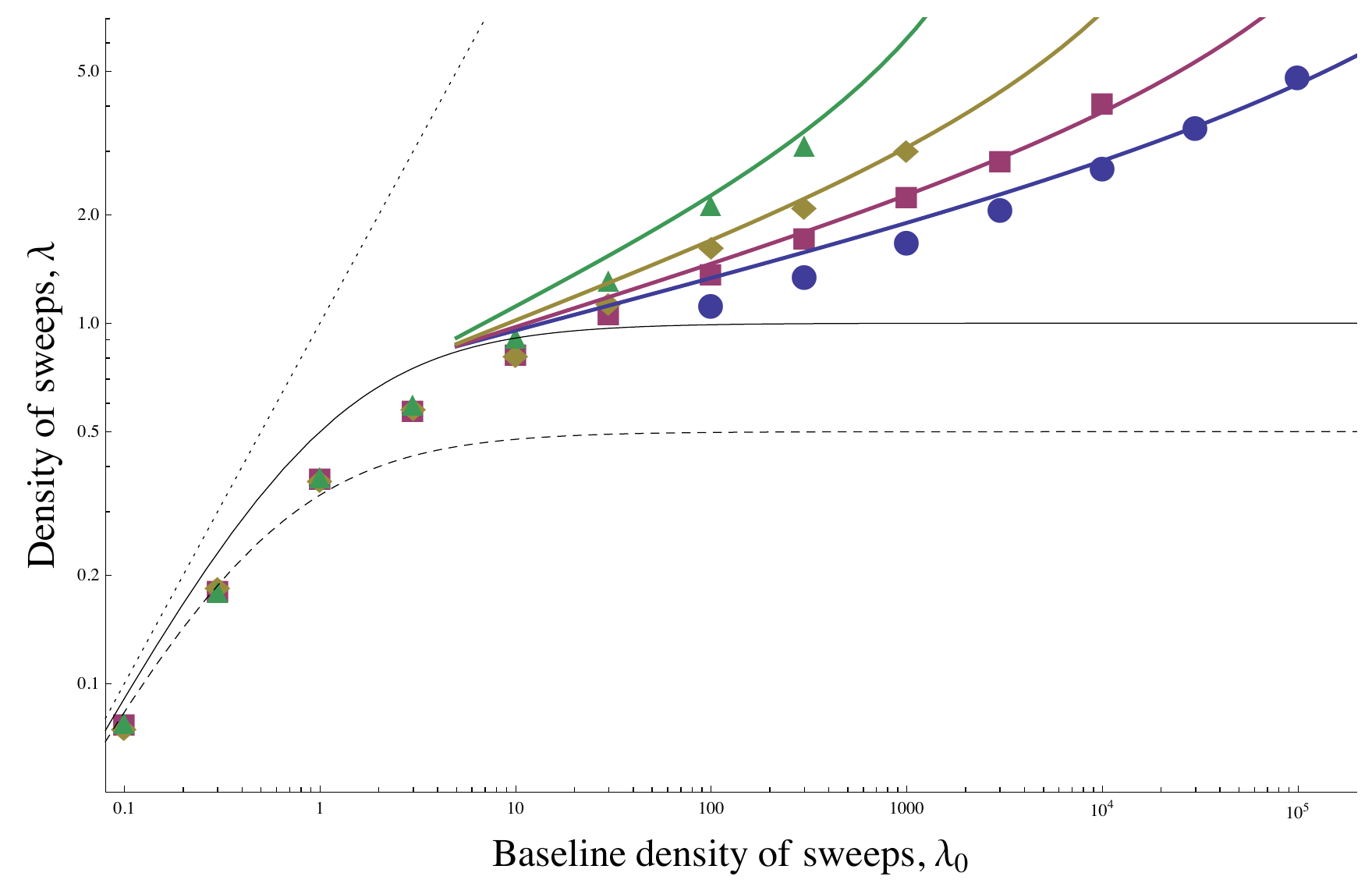}
 \caption{\label{fig:v}
 Rate of sweeps per unit genetic map length, $\lambda$, as a function of the rate in the absence of interference, $\lambda_0=2N\mu_b s$.
Symbols show simulation results for different fixed population sizes ($\log_{10}N=4,5,6,7$, moving down) with varying mutation rates $U_b$.
Colored curves show the matching analytical approximations given by \eq{lam},
which is accurate for strong interference ($\lambda_0 \gg 1$).
For weak-to-moderate interference ($\lambda_0 \lesssim 1$), $\lambda$ is described by \eq{lamweak} (solid black curve), which depends only on $\lambda_0$.
The dotted line shows $\lambda=\lambda_0$. 
The dashed black curve shows the prediction of \cite{weissman2012}; 
simulation results for $N\le 10^6$ with $\lambda_0 \le 3 \times 10^3$ are also from that paper.
$s=0.05, \, R=1$ for all points. Points are averages of $2-5 \times 10^3$ generations of steady-state evolution.
 }
 \end{figure}

To make this more precise, we will focus on the case in which the selective coefficients of beneficial mutations cluster tightly around a typical value $s$; 
we consider the case of exponentially-distributed effects below.
We will further focus on the case in which selection is strong relative to mutation, $s> \mu_b\rc$; we 
consider the biological plausibility of this assumption in the Discussion.
With these assumptions, we can apply the traveling-wave analysis of \cite{cohen2005a}, \cite{desai2007c}, and \cite{rouzine2008} to the evolution of each chunk
 to find $\rc$ and $\lambda$ self-consistently.
 We do this in \SI{calc}, and find that they are approximately given by
\begin{align}
\rc & \approx \frac{s}{\log\left[\log(\gamma)/\mu_b\right]},\label{rc}\\
\lambda & \approx \frac{2 \log( \gamma)}{\log\left[\log( \gamma)/\mu_b\right]}-1.\label{lam}
\end{align}
In order to cover a range of different reproduction models at once, we have written \eqs{rc}{lam}
 in terms of the scaled parameter $\gamma\equiv2Ns/\var$, 
where $\var$ is the variance in offspring number. 
($\var=2$ and $1$ in the Moran and Wright-Fisher models, respectively).
Simulations confirm that \eq{lam} accurately describes the rate of adaptation when interference is strong, $\lambda_0\equiv\mu_b \gamma \gg 1$ (\fig{v}).
Note, however, that some of the close match between the approximations and the simulations is due to lucky cancellations of error terms;
 see \SI{calc} and \SI{loose},
where we estimate the corrections due to interference among chunks.

For moderate interference, $\lambda_0 \sim 1$, we conduct a similar analysis in \SI{weak},
and find the expression
\begin{equation}
\lambda \approx \frac{\lambda_0}{1+\lambda_0}.\label{lamweak}
\end{equation}
This is similar to the approximation of \cite{weissman2012}, $\lambda\approx\lambda_0/(1+2\lambda_0)$,
but \fig{v} shows that it is less accurate for $\lambda<10$.
This is because interference among chunks has a significant effect in this regime;
see \SI{loose} and \SI{weak}.

To derive \eqs{rc}{lam}, we assumed that selection was stronger than mutation ($s>\mu_b \rc$).
\eq{rc} shows that this 
is consistent, given our earlier assumption that $\mu_b<1$.
In the opposite case, $\mu_b>1$, in which individual mutations are only weakly selected, 
\cite{neher2013} follow a very similar approach to find a result analogous to \eq{rc}, with $ \rc\sim s \sqrt{\mu_b/\log[\gamma \sqrt{\mu_b}]}$ in our notation.
\cite{good2013} also consider the genetic diversity in this case using a similar approach.
 It is surprising that the relative strength of mutation to selection 
 depends more on the frequency of recombination 
instead of the strength of selection.

\subsection*{Exponentially distributed mutational effects}

We now consider the case in which the effect of a beneficial mutation, $s$, rather than being a fixed value, is drawn from an exponential distribution with mean $\evs$.
We can take a similar approach as above, but now the evolution of the approximately asexual chunks is described by the analysis of \cite{good2012}, rather than \cite{desai2007c}.
The density of adaptive substitutions $\lambda$ is not a very useful quantity in this case, since the substitutions will have a 
range of effects, so we will instead focus on $v$.
Given $\rc$, the width and speed of the fitness wave of each chunk can be found using Eqs.~(13--14) from \cite{good2012}, 
which we reproduce in \SI{exp} using our notation. 
These values can then be plugged back into \eq{rcq} (with $s$ replaced by $\evs$)
to solve for $\rc$ self-consistently.

\begin{figure}
 \includegraphics[width=6in]{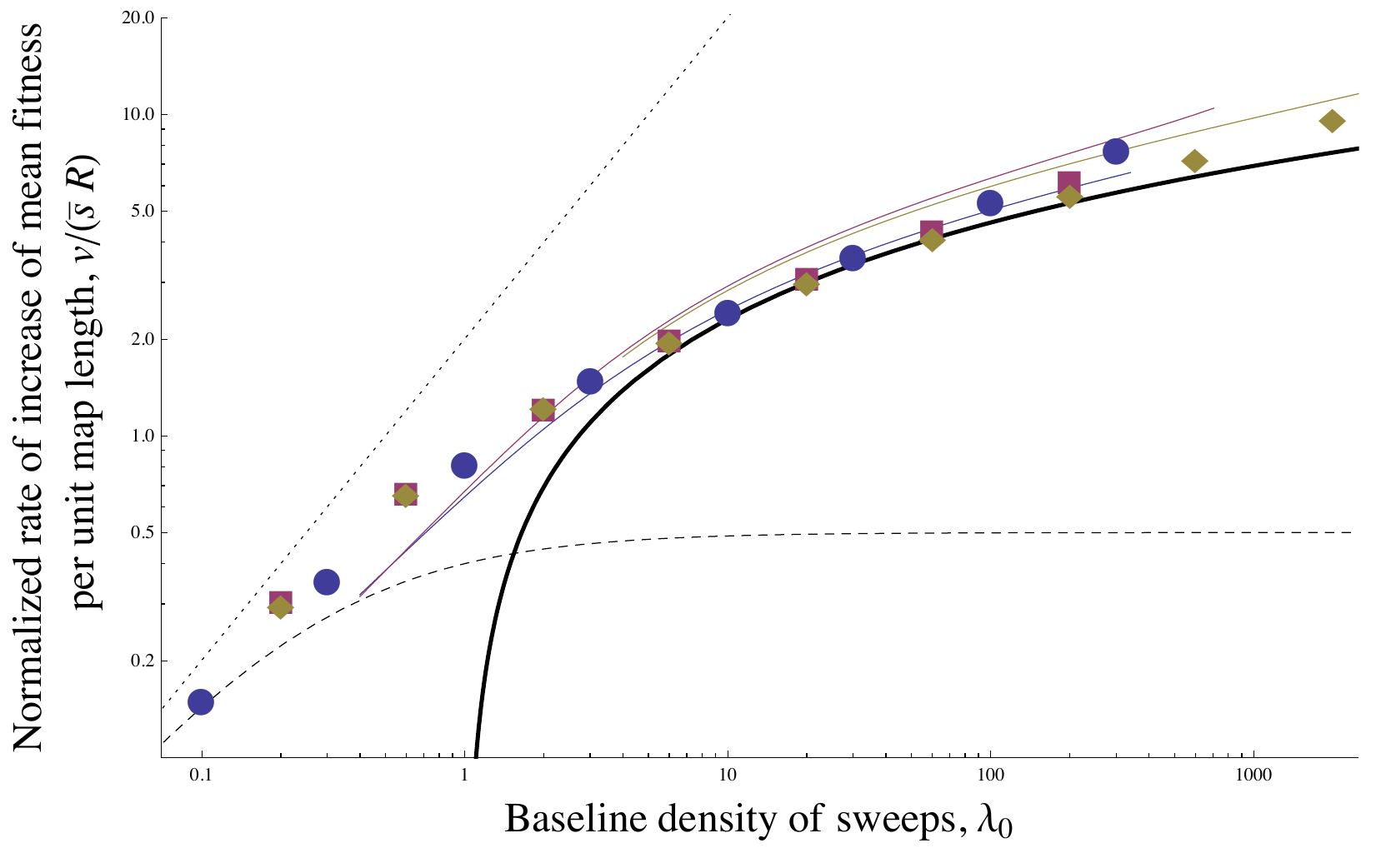}
 \caption{\label{fig:vexp}
 Normalized rate of fitness increase per unit genetic map length, $v/(\evs R)$, as a function of the density of sweeps in the absence of interference, $\lambda_0=2N\mu_b \evs$,
 for exponentially-distributed mutational effects with mean $\evs$.
Symbols show simulation results for different fixed population sizes and mean effects ($N=10^4$, $\evs=0.05$, blue; $N=10^5$, $\evs=0.01$, magenta;
$N=10^6$, $\evs=0.01$, gold) with varying mutation rates $U_b$.
The solid black curve is the analytical approximation given by \eq{vexp}, $v/(\evs R)\approx \log(\lambda_0)$;
it  is accurate for strong interference ($\lambda_0 \gg1$).
The thin colored curves show numerical solutions to Eqs.~(\ref{rcq}), (\ref{good13}), and (\ref{good14}).
The dotted line shows the value in the absence of interference, $v/(\evs R)=2\lambda_0$. 
The dashed black curve shows the approximation of \cite{weissman2012}; the simulation results in blue are based on data also used in that paper.
The beneficial mutation densities are approaching $\mu_b =1$ at the right-hand edge of the plot, 
so our analysis (which assumes $\mu_b\ll1$) is reaching the end of its validity.
Points are averages over $4-7\times 10^3$ generations of steady-state evolution.
 }
 \end{figure}

 In the strong-interference regime ($v\gg \evs R, \rc<\evs$), 
solving the system gives a simple approximate expression for the rate of adaptation (see \SI{exp}):
 \begin{equation}
v/(\evs R) \approx\log(\gamma \mu_b), \label{vexp}
\end{equation}
which matches well with simulation results (\fig{vexp}).
 It is interesting that \eq{vexp} is much simpler than the corresponding ones for both a population with fixed selective coefficients (\eq{lam}) and 
 for an asexual population with exponentially-distributed coefficients \citep{good2012}.

\begin{figure}
 \includegraphics[width=6in]{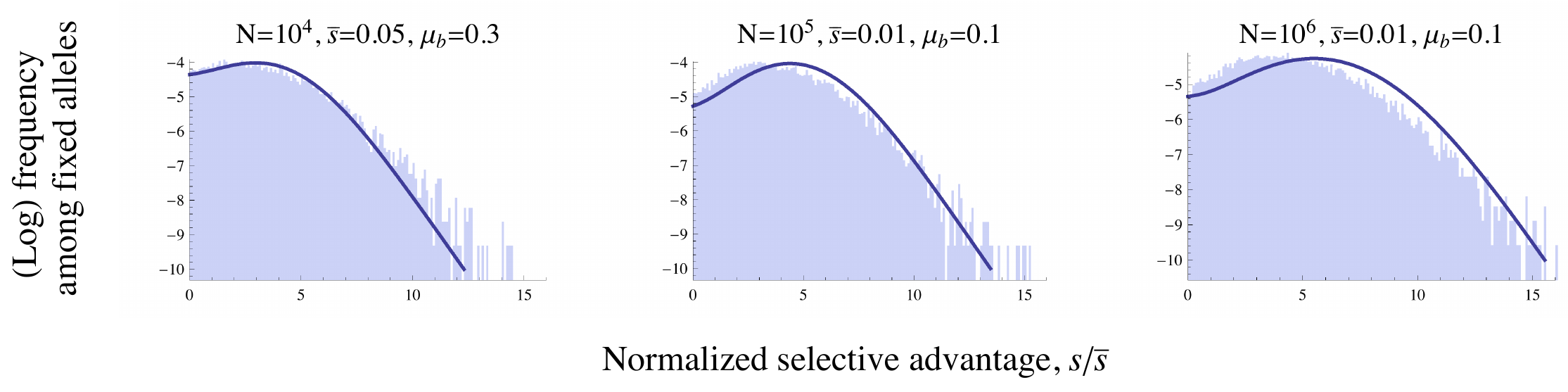}
 \caption{\label{fig:sdist}
Distribution of effects of fixed beneficial mutations. Histograms show simulation results; curves show numerical predictions.
Beneficial mutations are produced at rate $U_b$ with exponentially-distributed effects, with mean $\evs$.
The numerical approximations generally fit well, but underestimate the probability of fixation of weakly-selected mutations.
Histograms are the result of $4-7\times 10^3$ generations of steady-state evolution.
 }
 \end{figure}

Besides the rate of adaptation, we can also find the distribution of fixed effects of mutations, using Eq.~(11) of \cite{good2012}.
\fig{sdist} shows that this gives a rough match to simulation results, 
although the the probability of fixation of small-effect mutations is underestimated.
This may be due to inaccuracies in our approximations, or in the original asexual equation \citep{fisher2013}.
Both the analytical and simulation results indicate that 
over most of the simulated parameter range successful mutations generally have large selective 
coefficients but arise on average genetic backgrounds;
only at the highest simulated $\mu_b$ values (those shown in \fig{sdist}) do the backgrounds 
contribute significantly to the fitness of successful mutants.

\subsection*{Neutral diversity}

We want to check what neutral genetic diversity should look like in a population evolving under the dynamics described above.
The expected pairwise coalescence time at a neutral locus in a traveling fitness wave is approximately twice the time for the wave to go the distance from its mean to its nose \citep{desai2013},
i.e., $\approx 2/\rc$.
The neutral nucleotide diversity (the expected number of neutral differences between a random pair of genomes) should therefore be 
$\pi\approx 4U_n/\rc$, where $U_n$ is the neutral mutation rate.
Plugging in the value of $\rc$ from \eq{rc}, this is:
\begin{equation}
\pi \approx 4U_n\log\left[2\log(2\gamma)/\mu_b\right]/s.\label{pi}
\end{equation}
Given that $\pi$ is proportional to $N$ in a neutrally-evolving population, it may seem surprising that \eq{pi} only depends on the population
size very weakly, through a double logarithm. 
This is a consequence of the fact that the speed and length of the asexual traveling wave have nearly the same dependence on $N$,
so the coalescence time, which is given by their ratio, is nearly independent of $N$ \citep{desai2013}.

\begin{figure}
 \begin{center}
 \includegraphics[width=6in]{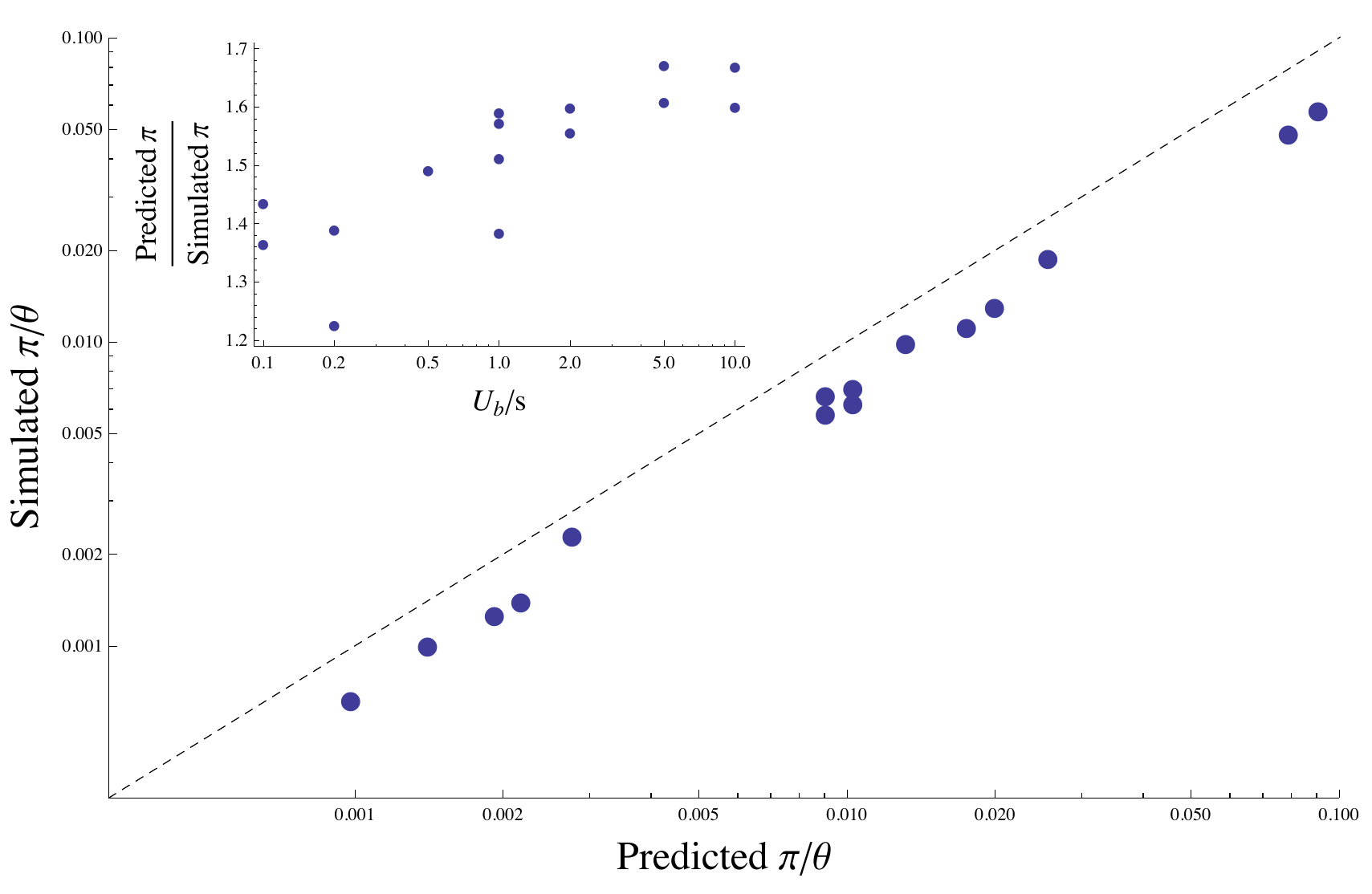}
 \end{center}
 \caption{\label{fig:pi}
Comparison between  simulated neutral nucleotide diversity $\pi$ 
and  the value predicted by \eq{pi}. 
$\pi$ is normalized by $\theta\equiv2NU_n$, where $U_n$ is the genomic neutral mutation rate.
The dashed line shows equality.
 \eq{pi} shows roughly the right parameter dependence, but consistently overestimates $\pi$ by $\approx 50\%$.
 Inset: the inaccuracy appears to increase for larger values of $U_b/s$.
Simulations use a range of parameter values, with $N=10^4$ and $10^5$; $s=0.01$, $0.05$, and $0.1$; and $\mu_b=0.01$, $0.05$, and $0.1$.
All points show the end results of $2 \times 10^4$ generations.}
 \end{figure}

From simulations, \eq{pi} appears to have the correct scaling with the parameters, but consistently overestimates $\pi$ by about $50\%$ (see \fig{pi}).
This difference is not so surprising, given that our derivation of $\rc$ was only approximate. 
The inset in \fig{pi} shows that \eq{pi}'s accuracy does appear to improve for low values of $U_b/s$,
although not by much.
Part of the inaccuracy may be because the pairwise coalescence time of a traveling wave is only equal to twice the nose-to-mean time
in the limit of very wide waves \citep{desai2013}; 
in our case, the width is $\approx \lambda +1$ (see \SI{calc}), which is never very large.

\begin{figure}
 \includegraphics[width=6in]{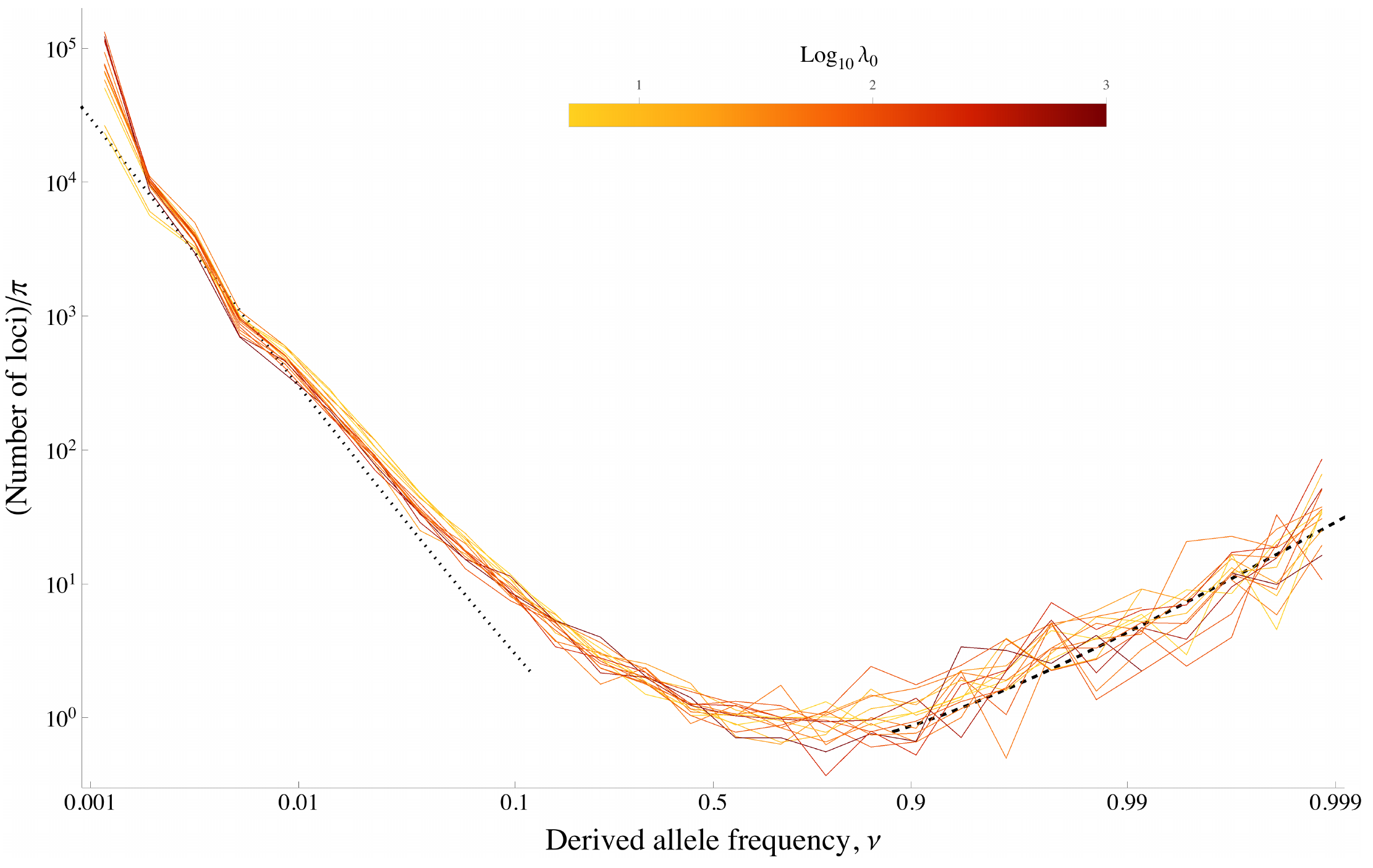}
 \caption{\label{fig:sfs}
 Neutral site frequency spectra, normalized by the neutral nucleotide diversity $\pi$.
 The dotted ($\propto \nu^{-2}$) and dashed  ($\propto \left[(\nu-1)\log(1-\nu)\right]^{-1}$) black lines show the shape expected under the Bolthausen-Sznitman coalescent
 for low  and high  frequency, respectively, with the constants of proportionality fitted to the data by eye.
The range of parameters is the same as \fig{pi}, and the curves are colored according to the baseline density of adaptive substitutions, $\lambda_0$.
 }
 \end{figure}

Going beyond just the nucleotide diversity, \fig{sfs} shows the full one-locus site frequency spectrum.
Very wide traveling waves are expected to approach a Bolthausen-Sznitman coalescent \citep{desai2013},
giving a site-frequency spectrum with the characteristic scaling $\propto \nu^{-2}$ as the  derived allele frequency $\nu$ approaches 0,
and $\propto \left[(\nu-1)\log(1-\nu)\right]^{-1}$ as $\nu$ approaches 1 \citep{neher2013a}.
Again, the chunk waves are never very wide, so we would not expect this to be a very good approximation in our case;
however, the scaling appears to be accurate, particularly for $\nu\to1$.
Surprisingly, for $\nu\to0$, the simulations with the lowest values of $\lambda$ (i.e., the narrowest chunk waves) 
are the closest to the Bolthausen-Sznitman scaling.

\begin{figure}
 \begin{center}
 \includegraphics[width=6in]{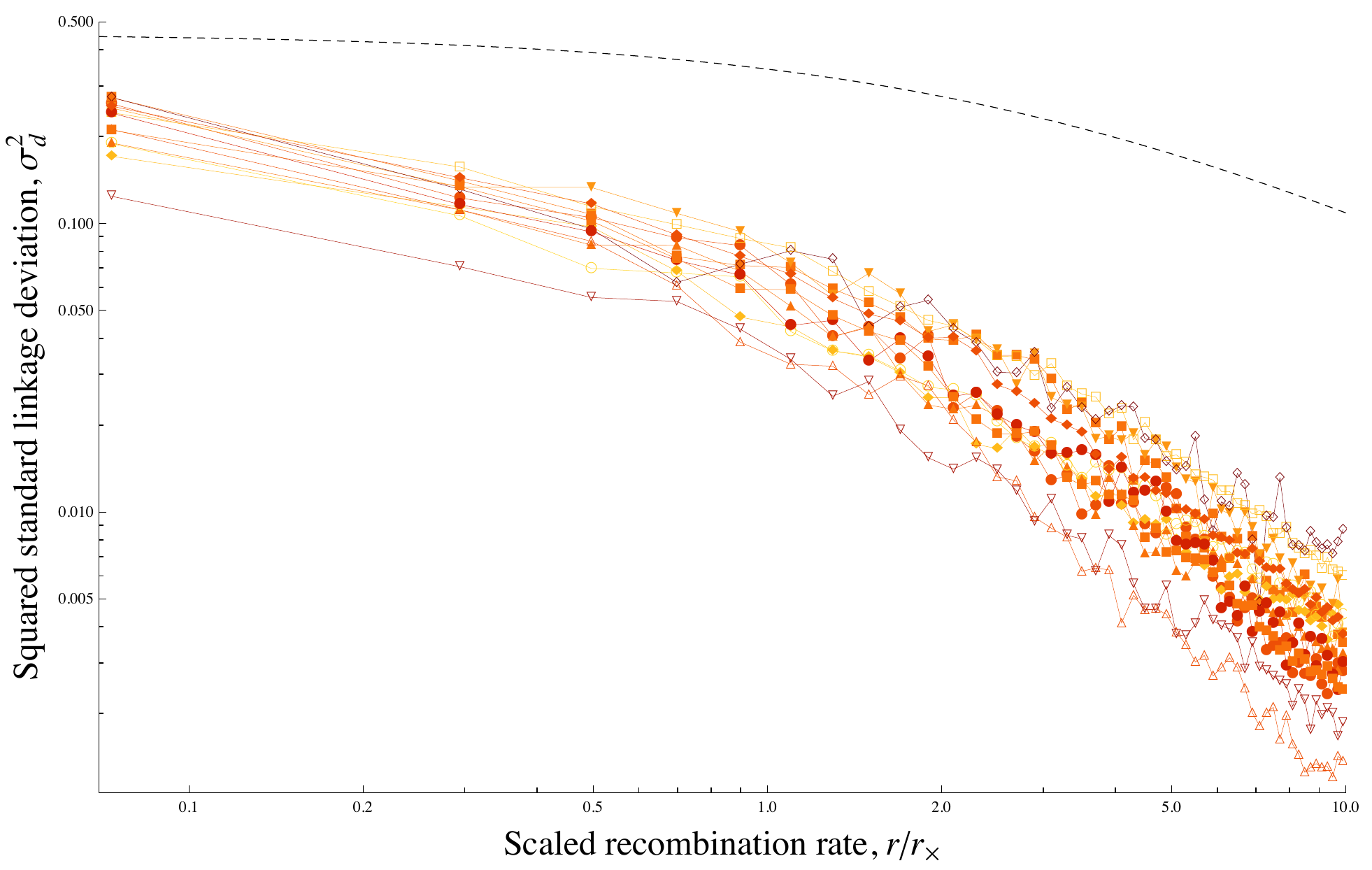}
 \end{center}
 \caption{\label{fig:ld}
The standard linkage deviation  ($\sigma_d^2$, \eq{sld}) between pairs of neutral mutations as a function of the recombination rate between them,
for a range of parameters.
 $r$ is normalized by the scale $\rc$ over which evolution is predicted to be effectively asexual (\eq{rc}).
 $\sigma_d^2$ falls off slowly for $r < \rc$, 
 and then rapidly for $r> \rc$.
 For the range of $r$ shown here, $\sigma_d^2$ is approximately  $\propto r^{-1/2}$ and $\propto r^{-2}$ for small and large $r/\rc$, respectively, 
 although it is expected to eventually approach maximum and minimum values.
 The dotted curve shows the expectation for a neutral Wright-Fisher population.
 The parameter range and coloring are the same as in \fig{sfs}.
 }
 \end{figure}

We also want to consider linkage disequilibria among loci.
Specifically, we look at the squared ``standard linkage deviation'', $\sigma_d^2$,
defined for a pair of loci with mutant allele frequencies $\nu_1, \nu_2$ and double-mutant haplotype frequency $\nu_{12}$ as \citep{ohta1969}
\begin{equation}
\sigma_d^2 \equiv \frac{\ev{(\nu_{12}-\nu_1\nu_2)^2}}{\ev{\nu_1(1-\nu_1)\nu_2(1-\nu_2)}}.\label{sld}
\end{equation}
(Compared to other measured of linkage disequilibrium, $\sigma_d^2$ has the advantage of being relatively insensitive to associations among rare alleles
and easy to calculate analytically.)
\fig{ld} shows that $\sigma_d^2$ between loci in simulations decays with the recombination fraction $r$ between them, 
but does so much more quickly for $r > \rc$ than for $r < \rc$, suggesting that $\rc$ is indeed an appropriate scale.
\fig{ld} also shows that the pattern of $\sigma_d^2$ is very different than that of a neutral Wright-Fisher population with size $N_e=\pi/(2U_n)$
(\cite{ohta1969}, Eq.~(18)).
This is to be expected, since the underlying coalescent process is also very different.
Note that while this contrasts with \cite{zeng2011}'s finding that the effect LD of background selection on deleterious mutations \textit{could} be accounted for 
by adjusting $N_e$ in this way, they only considered the regime in which the deleterious alleles are in linkage equilibrium with each other,
and it is not clear if this result extends to the regime in which the deleterious alleles interfere with each other (``weak selection Hill-Robertson interference,'' \cite{mcvean2000,kaiser2009}).

\section*{Simulation methods}

To check the accuracy of our approximations, we conducted individual-based simulations.
Simulated populations reproduced according to the Wright-Fisher model.
Individuals were obligately sexual (with selfing allowed), and each genome consisted of a single linear chromosome with uniform crossover.
The average rate of crossover was $R=1$ per genome per generation for all simulation data shown here.
There was a constant supply of beneficial mutations, and no back-mutations.
In the simulations used to study the site frequency spectrum and the dependence of linkage disequilibrium on recombination fraction, there was also
a constant supply of neutral mutations. 
For population sizes $N\le10^6$, the genome was modeled as continuous, with an (effectively) infinite number of loci.
For larger $N$, this required too much memory, and the genome was instead modeled as having 500 evenly-spaced loci, each with an infinite number of possible alleles.
The recombination fraction between adjacent loci in this model was small compared to the predicted chunk length $\rc$ for all simulated parameter values, 
with the ratio between the two reaching a maximum of $\approx .2$ for $\lambda_0=10^5$ in \fig{v}.
Even this discrete-locus model became very memory- and computation-intensive for large populations; 
$N=10^7$ was already pushing the limits of our hardware.

Simulating populations with exponentially-distributed mutational effects was particularly difficult.
The mean effect $\evs$ had to be kept small to avoid having a substantial fraction of fixed mutations with very large selective coefficients, $s\sim\bigo{1}$,
when interference was strong.
(Our approximations assume $s \ll 1$ throughout.)
\fig{sdist} shows that with $\evs=0.05$ our simulations were already beginning to approach this regime for $\lambda_0 \gtrsim 100$.
But for very small values of $\evs$, mutations took a long time to sweep through the population, increasing the memory usage of the simulations.
Even with $\evs=0.01$, population sizes of more than $N=10^6$ undergoing strong interference were computationally impractical.
Simulations were therefore limited to a fairly narrow range of parameters.

\section*{Discussion}

When biological adaptation is controlled by a combination of several
evolutionary forces with widely-varying strengths, it is important to have
simple order-of-magnitude estimates of which combinations of forces
are important and how. In this article, we have examined
well-mixed populations in which adaptation is driven by the 
interaction of beneficial mutation, recombination (via
crossover), selection, and genetic drift.
 We have hypothesized
and checked by explicit simulations that, restricted to a suitably
chosen local genomic scale, the dynamics of this sexual case
reduce to known results for asexual adaptation. 
The whole genome can be thought of as being
subdivided into uncorrelated chunks, each evolving
effectively asexually. We determined the
characteristic chunk length self-consistently by requiring that the
local coalescence times are just about long enough for neighboring
chunks to become decorrelated through recombination. Despite the
simplicity of our approximation, the resulting predictions for the
speed of adaptation, linkage disequilibrium, and the distribution of
fixed mutational effects compare well with the simulations.

\cite{fisher1930} noted that the potential increase in the
rate of adaptation of sexual populations over asexual ones is given by
``the number of different loci in the
sexual species, the genes in which are freely interchangeable in the course of descent'' (p.~123).
To understand when this is relevant to evolution,
 it is necessary to understand under what circumstances 
this maximum potential increase is approached, given a fixed number of recombining loci 
\citep{maynard-smith1971,kim2005}.
Here we have focused on another aspect of the problem,
 considering populations adapting at their maximum possible rate 
and  investigating  how  ``the course of descent'' and the 
number of ``freely interchangeable'' genes interact 
to determine each other.
\cite{park2013} take a hybrid approach,
considering two asexual loci, each experiencing strong clonal interference,
 and investigating how
frequently recombination between them needs to occur for them not to
interfere with each other.
They find that the rate of adaptation slowly increases over a broad range of recombination
rates ($>3$ orders of magnitude for some parameter combinations).
This suggests that in their model successful mutations at each locus have to occur in individuals
that are also highly fit at the other locus, with the required fitness slowly
decreasing with increasing recombination rate.
The dependence of the rate of adaptation on the rate of recombination is much 
weaker than the nearly linear relation found in our model.

We have already discussed the connections between our analysis and
the closely related work by \cite{weissman2012} and \cite{neher2013}, 
who examine the same question in the parameter regimes $\mu_b \lesssim 1/(Ns)$
and $\mu_b>1$, respectively.
Our analysis bridges the gap between these two analyses, focusing on the case 
$1/(Ns) < \mu_b < 1$ in which the
density of beneficial mutations $\mu_b=U_b/R$ 
 is large enough so that 
interference among them is strong, but not so large that it overwhelms the
effect of selection on individual mutations.
For large populations, this is a broad region of parameter space.
(Note that $N$ in the condition above is the short-term effective population
size, which may be many orders of magnitude larger than the long-term 
effective size $N_e$ measured from heterozygosity -- see \fig{pi}.)
Which natural populations might plausibly fall inside it?
Almost all obligatorily outcrossing organisms certainly satisfy the condition $\mu_b<1$, 
since they typically have \textit{total} mutation rates on the same
order as rates of crossover, and only a small (albeit often
unknown) fraction of those mutations are beneficial. 
However, many of them are likely to have sufficient recombination that
interference among beneficial mutations is negligible, $1/(Ns) > \mu_b$ 
\citep{weissman2012}.

Organisms with lower rates of outcrossing, such as viruses and
selfing and facultatively sexual eukaryotes are more likely 
candidates.
However, very little is known about natural rates of recombination for most
of these species. 
It is difficult to directly measure short-term recombination rates in natural conditions,
and rates inferred from diverged genomes measure some convolution of 
recombination and selection on recombinants, rather than recombination itself.

Bacteria, for which recombination occurs primarily via the exchange of
short stretches of DNA rather than crossovers, are not described by our model
(unless the rate of exchange at each site is large compared to the coalescence time,
which seems unlikely).
Instead, these populations are described by the analysis of \cite{neher2010}
and \cite{neher2011}, which assume that all loci have approximately the same
recombination probability with each other. 
Interference is thus similar to that among unlinked loci in our model (see \SI{loose}),
with the same parameter $4v/\oxf^2$ controlling the strength.
For organisms that primarily recombine via crossovers and also
have $4v/\oxf^2 \gg 1$, both forms of interference are likely to be important.

Of the organisms with limited outcrossing, HIV evolving within a host has perhaps 
the best-characterized natural mutation and recombination rates. \cite{neher2010a}
estimate that in chronic infections the recombination rate is
approximately $10^{-5}$ per base. Interestingly, just as in obligate sexuals, 
this is on the same order as the 
per base mutation rate \citep{abram2010}, implying that $\mu_b\ll1$.
The rate of substitutions is also on the same order, ranging up to
approximately $5\times 10^{-5}$ per base depending on the gene and stage of infection
 \citep{shankarappa1999}.
If a substantial fraction of the substitutions are adaptive (as must be
the case when the substitution rate exceeds the mutation rate),
the density of adaptive substitutions is high enough that HIV is in the 
strong-interference regime described by our model, with $1/(Ns) < \mu_b$.
If the selective advantages of the adaptive substitutions are on the order of
$1\%$ \citep{neher2010a}, our model predicts that the genome 
(with length $\approx 10\text{kb}$)
consists of tens of effectively asexual chunks with lengths of hundreds of base pairs.

The above back-of-the-envelope calculation should not be taken too seriously.
Our model leaves out many features that are likely
to be important to adaptive evolution, both of HIV and more generally. 
Most obviously, we ignore deleterious mutations, which are likely to make 
up the vast majority of all selected mutations. 
It is unclear if our results still apply when the majority of fitness variance is due
to deleterious mutations rather than sweeps.
We also ignore weakly-selected beneficial mutations.
If nearly all mutations are selected (with most being deleterious or weakly beneficial)
then the total selected mutation rate might be
on the same order as $R$ or even larger.
If sweeps are rare, then this situation is covered by \cite{neher2013}'s
approach, but if sweeps are also common in addition to background selection,
a combination of their method and that of this paper may be necessary.

In addition to deleterious and weakly beneficial mutations, we also ignore 
several other factors that are likely to be important. 
In large populations, adaptation may be driven by 
selection on standing variation due to environmental change rather 
than by new mutations \citep{hermisson2005}, in which case the amount of interference among sweeps
depends on how long the alleles were present in the population as neutral  
or deleterious variation.
Population structure can also affect the amount of interference, as it tends to slow
down selective sweeps, lowering the threshold rate 
 at which they begin to overlap and interfere \citep{martens2011}.
We also ignore epistasis, which can overwhelm recombination and 
preserve linkage disequilibrium if it is strong enough \citep{neher2009}.

In the light of our results, we may revisit Weismann's hypothesis that
sex is selected for because recombination reduces clonal interference and thus
speeds up adaptation. Our model generally supports this hypothesis, as
the speed of adaptation is predicted to be roughly proportional to
$R$, as can be seen from \eq{lam}. However, we have not investigated 
whether this can effectively select for a modifier allele increasing recombination.
Note also that the speedup due to sex
only arises if the total map length $R$ is larger than the characteristic
chunk size $\rc$; otherwise the whole genome is effectively
asexual and recombination is too rare to have a significant effect. 
If we consider a facultative sexual such as yeast, is it plausible that 
recombination is frequent enough to substantially speed up adaptation?
Assuming $s\sim 1\%$ and $U_b\sim10^{-5}$, as \cite{desai2007a}
measured for \textit{S.~cerevisiae} in a laboratory setting, \eq{rc} gives a minimum value of 
$R\sim 2\times 10^{-3}$, roughly independent of $N$. 
Given that \textit{S.~cerevisiae} undergoes $\approx 43$
crossovers per mating, this corresponds to a minimum frequency of sexual
reproduction of $\sim 4\times10^{-5}$. Thus, even small, difficult-to-measure
rates of sex may be effective in alleviating Hill-Robertson
interference.

It is somewhat surprising that our mean field approximation based on a
typical block length worked, as it does not take into account
fluctuations in the chunk lengths. This may be in part due to a negative
feedback: if an anomalously strong clone arises at one location, it
leads to a larger linkage block. This in turn will increase the
interference among local sweeps, thus reducing the local density of
beneficial sweeps. Overall, these effects tend to push block sizes
towards a mean block size, as assumed in our argument. Note however,
that for broad distributions of fitness effects we observe
significant deviations from our simple predictions.

\section*{Acknowledgments}

We thank Nick Barton and Richard Neher for helpful discussions. 
DBW received financial  support from ERC grant 250152.
O.H. received financial support from the German Research Foundation (DFG), within the Priority Programme 1590 ``Probabilistic Structures in Evolution'', Grant no.~HA 5163/2-1.

\bibliographystyle{genetics}
\bibliography{references}

\clearpage

\begin{table}[!ht]
\caption{
\bf{Symbol definitions}}
\begin{tabular}{|c|l|}
\hline
\textbf{Symbol} & \textbf{Definition}\\
\hline
$U_b$ & Genomic beneficial mutation rate\\
$R$ & Total genetic map length (in Morgans); $R=1$ in all simulations\\
$\mu_b$ & Beneficial mutation rate per Morgan, $\mu_b\equiv U_b/R$\\
$N$ & Haploid population size\\
$s$ & Selective advantage of beneficial mutations\\
$\gamma$ & $2Ns$ divided by the variance in offspring number\\
$\rc$ & Genetic distance within which genomes are effectively asexual\\ 
$v$ & Rate of increase of population mean fitness\\
$\lambda$ & Rate of fixation of beneficial mutations per Morgan\\
$\lambda_0, v_0$ & Values of $\lambda$ and $v$ in the absence of interference\\
\hline
\end{tabular}
\begin{flushleft} The definitions of the main symbols used in the text. $U_b$, $R$, $\mu_b$, $N$, $s$, $\lambda_0$, and $v_0$ are population parameters, 
while $\lambda$, $v$, and $\rc$ are dynamical variables. 
\end{flushleft}
\label{tab:defs}
 \end{table}

   \clearpage
   
 \clearpage
 
\clearpage

  \clearpage
 
\clearpage

  \clearpage
 
\clearpage

\appendix

\section{Solving for $\rc$ and $\lambda$}\label{si:calc}

In this Appendix, we show how to use asexual traveling wave theory to determine the density of adaptive substitutions $\lambda$ 
and the length $\rc$ of the effectively asexual chunks.
We will follow \cite{desai2007c}'s intuitive argument for simplicity;
the same results can be derived from their formal calculations or from those of \cite{rouzine2008}.
For each chunk, define $\qc$ as the difference between the number of beneficial mutations in the fittest genotype 
in the population and the number in the average genotype. 
Typically, this fittest genotype will be present in no more than a few copies, and will most likely be lost to stochastic drift. 
However, there will usually be a genotype with $\qc-1$ mutations that has established and is starting to sweep through the population.
From the time when the genotype first establishes to when it reaches high frequency (i.e., the nose-to-mean time of the fitness wave)
 takes $\tnm\sim 2\log(\qc \gamma)/(\qc s)$ generations
 \citep{desai2007c}. (The factor of two comes from the fact that its mean selective advantage drops from $\approx(\qc-1)s$ to $\approx s$ over the course of the sweep.)
The length of the chunk of genome that stays linked over the course of the sweep is $\sim 1/\tnm$;
this sets $\rc$:
\begin{equation}
\rc\approx\frac{\qc s}{2\log{\qc \gamma}}.\label{rcq}
\end{equation}

Since $\qc-1$ mutations fix every $1/\rc$ generations in every chunk of genome of length $\rc$, the density of adaptive substitutions is simply
\begin{equation}
\lambda = \qc -1,
\end{equation}
and we can write everything in terms of $\lambda$, the quantity we want to find.
Doing this, our expression for $\rc$ is 
\begin{equation}
\rc \approx \frac{(\lambda+1) s}{2\log[ \gamma(\lambda+1)]}.\label{rcswp}
\end{equation}

\begin{figure}
 \includegraphics[width=6in]{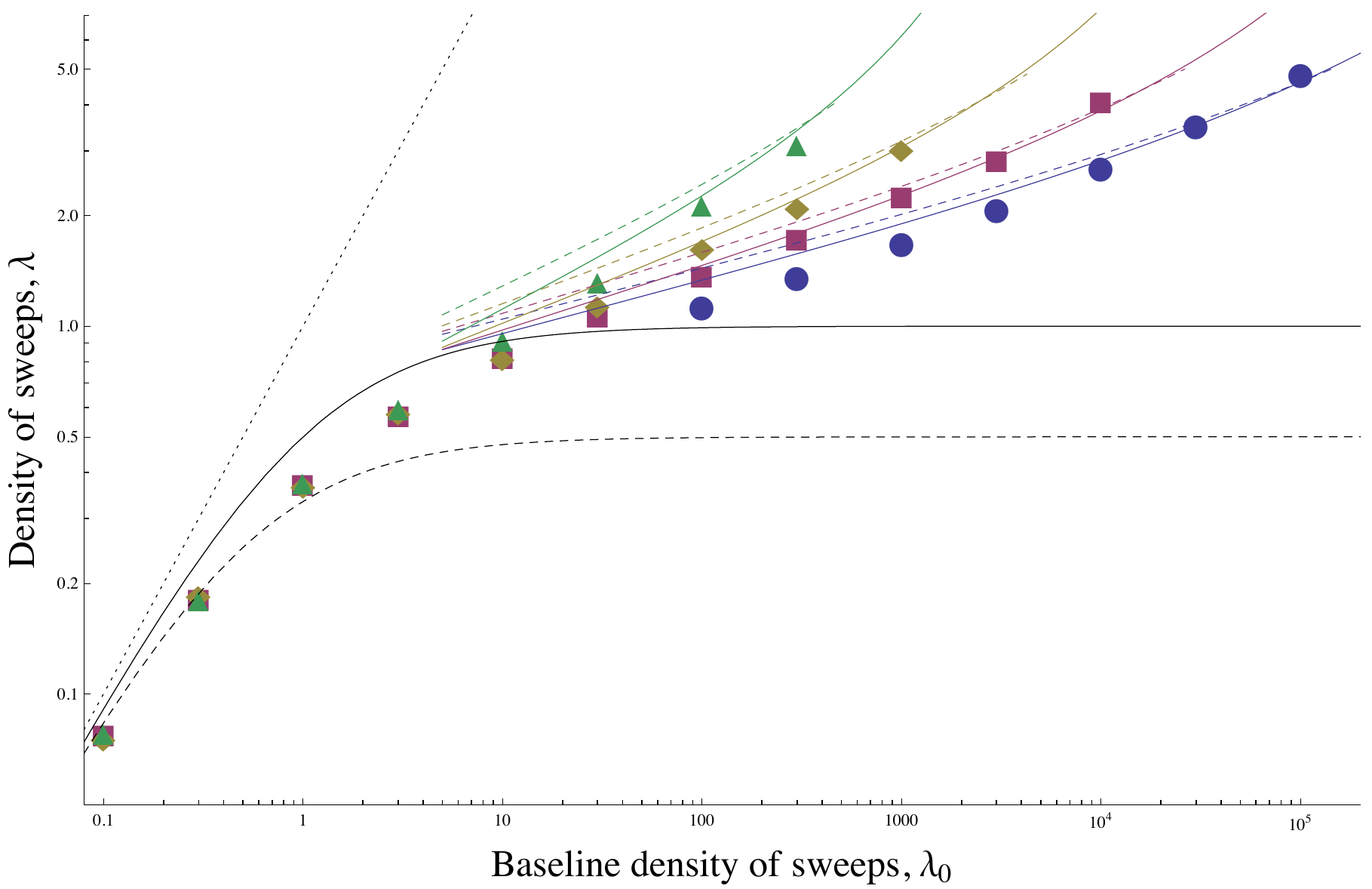}
 \caption{\label{fig:app1LL0}
 Rate of sweeps per unit genetic map length, $\lambda$, as a function of the rate in the absence of interference, $\lambda_0=2N\mu_b s$.
All points and curves are as in \fig{v}.
The additional dashed curves show the numerical solutions to \eqs{rcswp}{rcmut}, 
which are very close to the analytical approximations for large $\lambda_0$.
The analytical approximations are actually slightly closer to the simulation results, 
due to  cancellations with the errors introduced by ignoring the effects of loosely-linked loci (see \SI{loose}).
 }
 \end{figure}

To determine $\lambda$ and $\rc$, we can use the additional fact that in order to maintain
a steady wave in fitness space, in each chunk new mutations must be establishing at the same rate that 
 they are fixing, $\lambda \rc$. 
 In other words, the chunk with $\lambda$ mutations should produce an established chunk with an additional mutation
 in $t_1\sim1/(\lambda \rc)$ generations.
The number of copies of the chunk with $\lambda$ mutations $t<t_1$ generations after it 
establishes is  $n_\lambda(t)\approx\exp(\lambda s t)/((\lambda+1)s)$.
The total number of mutant genotypes it produces is $\approx \mu_b \rc \int_0^{1/(\lambda \rc)}dt\,  n_\lambda(t)$.
Each of these mutants has a probability $\sim (\lambda+1) s$ of establishing, so to have one successful mutant we must
have $\mu_b \rc (\lambda+1)s \int_0^{1/(\lambda \rc)}dt\,  n_\lambda(t)\approx 1$.
Evaluating the integral, we find the following condition:
 \begin{equation}
 \rc \approx \frac{s}{\log[\lambda s/(\mu_b\rc)]}.\label{rcmut}
 \end{equation}
\eqs{rcswp}{rcmut} can be rearranged to give 
\begin{align*}
\lambda & \approx \frac{2\log[\gamma(\lambda+1)]}{\log\left[\frac{\log[\gamma(\lambda+1)]}{\mu}\frac{2\lambda}{\lambda+1}\right]}-1\\
\rc & \approx s\left/\log\left[\frac{\log[\gamma(\lambda+1)]}{\mu}\frac{2\lambda}{\lambda+1}\right]\right..
\end{align*}
Expanding about $\lambda\approx 1$ and dropping $\bigo{1}$ terms in large logarithms gives \eqs{rc}{lam} in the main text.
The difference between the numerical solution to \eqs{rcswp}{rcmut} and the analytical approximation is negligible for strong interference; see \fig{app1LL0}

\clearpage

\section{Effect of background genetic variation at $r>\rc$}\label{si:loose}
  \setcounter{equation}{0}

We have assumed that each chunk of genome evolves roughly independently.
To confirm this, we need to understand how selection on variation at $r>\rc$ changes \eqs{rcswp}{rcmut}, i.e., 
how the rest of the genome affects the dynamics of a chunk that starts in the nose of the fitness distribution and sweeps to fixation.
We will just do a rough, approximate analysis, but even this gets quite involved.

As a first step, we will rewrite \eqs{rcswp}{rcmut} in a form that makes it clearer how they might be affected by variation in backgrounds:
\begin{align}
1/\rc & \approx \ev{\frac{\log[N/\nest]}{s_\text{sweep}}}\label{bgrcswp}\\
1/(\lambda \rc) & \approx \ev{\frac{\log[s_1/(\mu_b\rc\nest \bar p)]}{s_1}}\label{bgrcmut}
\end{align}
Here, $s_\text{sweep}$ is the mean selective advantage of genomes carrying the focal  mutation over the course of the sweep,
 $s_1$ is their mean selective advantage up to the point where they produce an additional successful mutation,
 $\bar p$ is the probability that a given additional mutation successfully establishes.
$\nest$ is the establishment size, which is roughly the number of copies of 
a successful chunk after $\sim \test=1/((\lambda+1)s)$ generations.
(See \cite{desai2007c} for a detailed discussion of the establishment dynamics.) 
All of these quantities are potentially affected by background genetic variation,
but we will see that $\bar p$ and $\nest$, the ones that depend on the shortest time scale $\test$,
that are the most affected.

It will be helpful to divide the genome into the region immediately around the chunk and the rest, 
and consider these two sets of loci separately.
Tightly-linked loci close to the chunk 
can stay linked for an extended period of time, and can be seen as perturbing the chunk's mean fitness. 
Variation at more distant loci is rapidly shuffled by recombination, and effectively increases the variance in offspring number in a way that 
is uncorrelated over the time scales relevant for selection and mutation.
To find the strength of these effects, we will first note that the density of variance in log fitness over the chromosome is $v/R=\lambda s$, 
so the standard deviation in log fitness due to
loci at a recombination fraction $<\lesssim r$ from the focal locus is $\sigma(r) = \sqrt{2\lambda s r}$
 (until $r$ saturates at $f/2$ for unlinked loci).

\subsection*{Tightly-linked loci}

First, consider the effect of the fitness of the initial chunk genotype at loci not too far away from the focal chunk.
For these to have a significant effect on average, their effect on fitness needs to be at least comparable to 
the chunk's selective coefficient. 
In other words, if $\sigma(r)< \lambda s$, then the region within $r$ typically does not contain 
enough variation to affect the dynamics.
Thus, we need to average over a region of width at least $\sim \lambda s$. 
This is also approximately the maximum scale over loci are effectively tightly-linked, 
as it is both roughly the region which stays linked over the shortest  relevant time scale $\test$,
and the region over which selection is strong enough relative to 
recombination to maintain unusually fit combinations,
i.e., $\sigma(r)\gg r$ for $r\ll\lambda s$, and $\sigma(r)\ll r$ for $r\gg\lambda s$
\citep{neher2013b}.

\begin{figure}
 \includegraphics[width=6in]{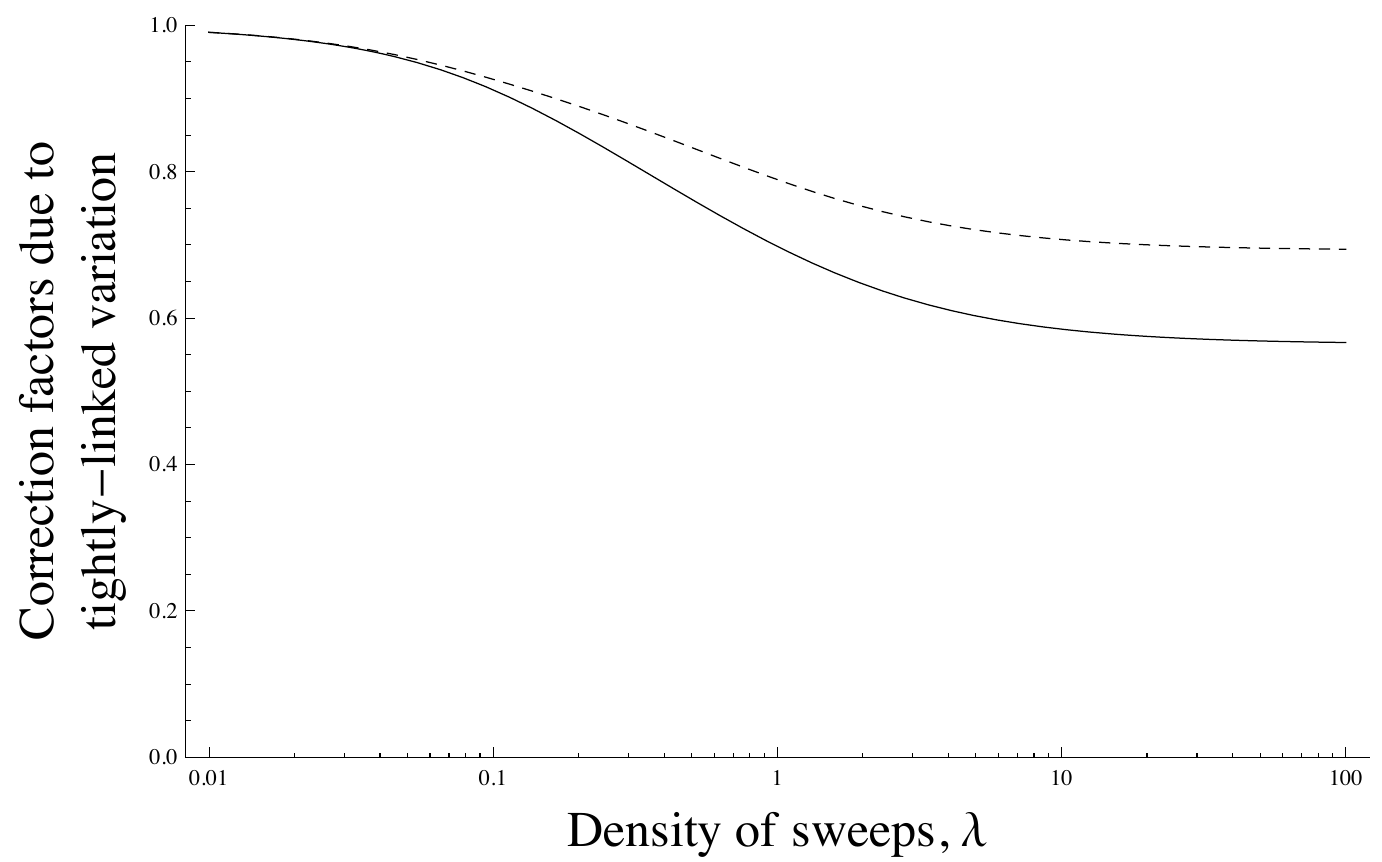}
 \caption{\label{fig:tight}
Corrections due to the effect of tightly-linked loci 
to the probability of establishment $\bar p$ of a new chunk at the nose (solid) and the typical number of copies
at establishment $\nest$ (dashed), as a function of the density of sweeps $\lambda$.
The correction factor for $\nest$ is shown as $\exp(\delta)$, where $\delta$ is the difference
due to tightly-linked loci in $\ev{\log(\nest)}$, so the similar shape of the two curves indicates that the corrections
largely cancel in \eq{bgrcmut}.
The corrections are never very large.
 }
 \end{figure}

Thus, the tightly-linked background will usually have a combined selective coefficient on the
same order as the chunk's own. 
More precisely, during the time $\test$ over which a new chunk with $\lambda +1$
mutations will either establish or go extinct, it will be strongly associated with a
particular genetic background with log relative fitness drawn from a normal distribution 
with standard deviation $\sigma(1/\test)\approx \sqrt{2\lambda(\lambda+1)} s$. This distribution
has initial  mean $\approx-\lambda s^2\approx0$ (\cite{weissman2012}, Supplementary Text 2),
dropping to $\approx-2\lambda s$ over the time $\test$ due to the increase in the population
mean fitness. 
Assuming that the probability of establishment is proportional to the average mean fitness of the 
chunk, averaging over this distribution of backgrounds
reduces the mean probability of establishment  by $ 30-40\%$ for
$\lambda\gtrsim1$ (\fig{tight}).

The dynamics of establishment also enter into \eqs{bgrcswp}{bgrcmut} through the mean log 
establishment size $\nest$.
Averaging over possible backgrounds weighted by their probability of establishing,
we find that $\ev{\log(\nest)}$ increased by $\approx \log(1.3--1.4)$ for $\lambda\gtrsim1$ (\fig{tight}).
Both this correction and the one to $\bar p$ have very small effects on the results.
In fact, in \eq{bgrcmut}, the two corrections approximately cancel each other: 
each mutant's probability of establishment is lower, so the lineage must produce more mutants
before one is successful,
but because the lineage is larger, it naturally does so.
(Note that the background variation at $r\sim (\lambda+1) s$ that determines $\nest$ has
largely been lost by the time the lineage starts producing new mutants, so it does not affect 
these new mutants' $\bar p$.)
In \eq{bgrcswp}, the increase in $\ev{\log(\nest)}$ can be included by adjusting $\gamma$.

Note that these tightly-linked loci do not have a significant effect on time scales long
compared to $\test$:
associations with initially poor or average backgrounds are reduced by recombination
to a small enough region of the genome that $\sigma(r)$ is small compared to the 
chunk's selective advantage, and associations with unusually good backgrounds 
are washed out as the population mean fitness catches up.
Thus $s_\text{sweep}$ in \eq{bgrcswp} and $s_1$ in \eq{bgrcmut}
are only slightly affected.

\subsection*{Loosely-linked loci}

We now consider the effect of loosely-linked loci at $r>(\lambda +1)s$.
In this case, we can use a generalization of an heuristic argument originally due to \cite{robertson1961} 
to obtain the amount of interference to a focal locus from a locus separated by a recombination fraction $r$ with log fitness variance $\sigma^2$, 
with $r\gg \sigma$.
An allele at the focal locus will typically have its log fitness perturbed by an amount $\sim\sigma$ due 
to variation at the interfering locus.
If each generation, each allele moved to a random background, the introduced variance in log fitness would just be $\sigma^2$
(and the additional variance in offspring number would be $\exp(\sigma^2)$).
 But the LD between the loci decays at a finite rate $r$, so a perturbation of size $\sigma$ will 
 still have an average residual effect $\approx \sigma \exp(-rt)$ after $t$ generations.
The accumulated perturbation after $t$ generations is therefore $\approx \int_0^t dt' \, \sigma\exp(-rt') = \sigma (1-\exp(-rt))/r$.
The accumulated variance in log fitness is the square, $\sigma^2(1-\exp(-rt))^2/r^2$. (Note that the variance in log fitness is highly correlated 
on time scales small compared to $1/r$, so the integration over time comes before squaring.)
As $t$ becomes large, this approaches $\sigma^2/r^2$, Robertson's result.
Applying this to unlinked loci in our model, we see that they increase the variance in offspring number by a factor $\exp(4v/\oxf^2)$, 
which is negligible for most of the simulated parameters.

Loosely-linked loci have a somewhat larger effect. 
Summing over all loci at $r>\lambda s$ and using that the density of log fitness variance is $\lambda s$,  the total effect is
\begin{equation*}
2\int_{(\lambda+1)s}^\infty dr\,  \frac{\lambda s}{r^2}  \approx \frac{1.5\lambda}{\lambda+1}.
\end{equation*}

\subsection*{Combined effect and limitations of results}

Combining the effects of tightly-linked, loosely-linked, and unlinked loci found above, 
the total effect of the background variation can be accounted for by adjusting the definition of $\gamma$:
\begin{equation}
\gamma = \gamma_0\left(0.7\exp\left[-\frac{1.5\lambda}{\lambda+1}-\frac{4v}{\oxf^2}\right]\right),\label{gamma}
\end{equation}
where $\gamma_0$ is the baseline value, $\gamma_0=2Ns/\var$.
The term $-4v/\oxf^2$ is the only way in which our results depend directly on $\oxf$, rather than $R$; 
it is negligible over almost all of the simulated parameter space.
(When it becomes large, the dynamics become sensitive to the details of the organism's life cycle; see \cite{weissman2012}.)
The initial factor of $0.7$  in \eq{gamma} is included assuming that $\lambda\gtrsim1$; for lower $\lambda$, it approaches 1.
It also has only a small effect for the regime $\lambda \gtrsim1$ where it applies.

\begin{figure}
 \includegraphics[width=6in]{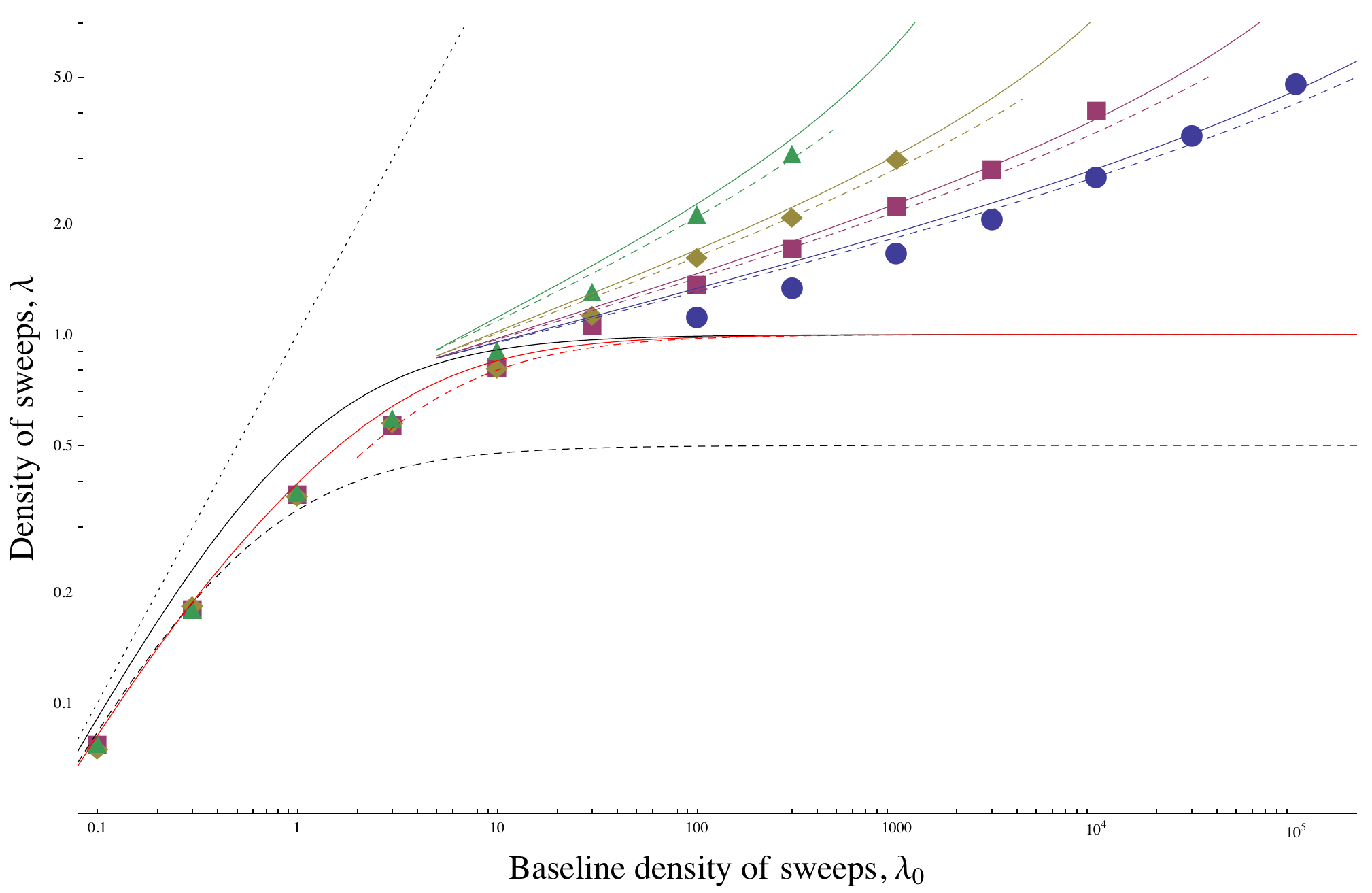}
 \caption{\label{fig:app2LL0}
 Rate of sweeps per unit genetic map length, $\lambda$, as a function of the rate in the absence of interference, $\lambda_0=2N\mu_b s$.
All points and curves are as in \fig{v}.
The additional dashed curves show the numerical solutions to \eqs{rcswp}{rcmut} with 
$\gamma$ given by \eq{gamma}.
The numerical and analytical curves are very close.
The solid red curve shows the closed-form prediction for moderate interference including the correction for loosely-linked loci, \eq{lamweakexact};
here, there is a noticeable difference, although it is still quite small.
The dashed red curve includes the additional  correction for tightly-linked loci, which makes only a negligible difference.
 }
 \end{figure}

In fact, over most of the parameter regime of interest, even the factor $\exp\left[-1.5\lambda/(\lambda+1)\right]$ has only a small effect on the results;
 see \fig{app2LL0}.
This is because generally either $\lambda$ is small (so $\gamma\approx\gamma_0$) or there is only a weak dependence on $\gamma$.
We include the correction in all numerical calculations, but omit it in all analytic approximations.
(It is of the same order as other small, omitted terms.)
The only region of parameter space where the effect is noticeable is
 for $\gamma\mu_b\approx1$, where $\lambda\sim1$ but is still fairly sensitive to $\gamma$.
We consider this regime of  moderate interference, $\lambda_0 \sim 1$, in \SI{weak} below.
Even in this case, leaving out the interference from loosely-linked loci would only increase the 
predicted $\lambda$ by about a factor of two.

We calculated \eq{gamma} by splitting the genome into the regions
$r<(\lambda+1)s$ and $r>(\lambda+1)s$, and analyzing these assuming tight and loose linkage with the focal chunk, respectively.
However, for both regions the dominant effect comes from loci with $r ~(\lambda+1)s$, 
where neither approximation is very accurate.
The exact form is therefore likely to be wrong,
but in any case our analysis shows that the error in \eqs{rc}{lam} from ignoring interference among chunks
are small, especially compared 
to the errors introduced by applying our idealized model to any real population.

\clearpage

\section{Moderate interference}\label{si:weak}
  \setcounter{equation}{0}

\eqs{rc}{lam} for $\rc$ and $\lambda$ are based on the analysis of a stable fitness wave, which is valid for $\qc\gg1$. 
For $\qc\approx 2$, the wave is always fluctuating, and this analysis is invalid. 
\cite{desai2007c} also investigate this case, and find that the time for a new mutation to arise and fix
 is dominated by the waiting time for a mutation to establish on a good genetic background, 
 which is given their Eq.~(46):
\begin{equation}
t_2 \approx \frac{1}{s}\log\left[ \gamma\left(\exp\left(\frac{s}{\gamma\mu_b\rc}\right)-1\right)\right].\label{desai46}
\end{equation}
$1/t_2$ is the rate at which mutations fix in each chunk, $1/t_2 = \lambda\rc$.
$\rc$ is still set by the inverse of the sweep time, which is now just the standard $t_\text{sweep}=\log(\gamma)/s$.
(\cite{desai2007c} find that the initial boost from arising on a good background makes little difference, 
as that background is itself rapidly approaching fixation.)

Plugging $\rc=s/\log(\gamma)$ into \eq{desai46} gives
\[
st_2 \approx \left(1+\frac{1}{\gamma\mu_b}\right)\log\gamma
\]
for $\gamma \gg 1$ and $\gamma\mu_b\approx 1$, 
the regime we are considering.
The density of substitutions is therefore 
\begin{equation}
\lambda \approx 1/(\rc t_2) \approx \frac{\gamma \mu_b}{1+\gamma\mu_b}.\label{lweakgm}
\end{equation}

As mentioned in \SI{loose}, the effect of loosely-linked loci outside the chunk must be take into account in this case
via \eq{gamma}.
(The effect of truly unlinked loci, with recombination fraction $\oxf/2$, is still negligible, as is the initial factor of $0.7$ due to tightly-linked loci.)
Substituting this into \eq{lweakgm}, we find the implicit equation
\[
\lambda \approx \frac{\lambda_0}{\exp\left[\frac{1.5\lambda}{\lambda+1}\right]+\lambda_0},
\]
the solution of which is well-approximated by 
\begin{equation}
\lambda \approx \frac{\lambda_0}{\exp\left[\frac{1.5\lambda_0}{2.5\lambda_0+1}\right]+\lambda_0}.\label{lamweakexact}
\end{equation}

\section{Exponentially-distributed effects}\label{si:exp}
  \setcounter{equation}{0}

Given the length $\rc$ of an effectively asexual chunk of genome,
we can characterize its evolution by
$\qc$, defined in this case as the typical relative fitness of the fittest chunk genotype $\evs$,
and $\vc=v\rc/R$, the rate at which the chunk's mean fitness is increasing. 
$\qc$ and $\vc$ can be found as functions of $\rc$ using Eqs.~(13--14) of \cite{good2012},
which are, in our notation:
 \begin{align}
2 & = \mu_b\rc\sqrt{\frac{2\pi}{\vc}}\left(1+\frac{1}{\qc}+\frac{\vc}{\qc\evs^2-\vc}\right)\exp\left[\frac{(\qc\evs-\vc/\evs)^2}{2\vc}\right] \label{good13}\\
1 & = \frac{\gamma\mu_b\rc\evs}{\vc}\left(\qc^2-\vc/\evs^2+2\qc+2\right)\exp\left[-\qc-\frac{\vc}{2\evs^2}\right]\label{good14},
 \end{align}
 where $\gamma \equiv 2N\evs/\var$. 
The solution to these equations can then be used with \eq{rcq} to determine $\rc$.
For $\qc\gg 1$, the rate of advance can be written simply as $\vc \approx \qc \rc\evs$, 
giving the very rough but simple solution
\[
\qc \approx \log(\gamma \mu_b),
\]
from which $\rc$ and $\vc$ follow directly.
This form can be guessed just from \eq{good14} -- the large factor of $\gamma \mu_b$ out front must be roughly canceled by the exponential factor $\exp(-\qc)$.

\end{document}